\documentclass[times,multirow,preprint,sort&compress,times]{elsarticle}
\usepackage[utf8x]{inputenc}
\usepackage{graphicx,epsfig,color,booktabs}
\usepackage{amssymb,amsmath,a4wide,caption}
\begin{document}

\title{Scaling features in the spreading of COVID-19}

\author[1]{Ming Li}
\ead{minglichn@ustc.edu.cn}

\author[1]{Jie Chen}

\author[2]{Youjin Deng\corref{cor1}}
\ead{yjdeng@ustc.edu.cn}

\cortext[cor1]{Corresponding author}

\address[1]{Department of Thermal Science and Energy Engineering, University of Science and Technology of China, Hefei 230026, P. R. China}
\address[2]{Hefei National Laboratory for Physical Sciences at Microscale, Department of Modern Physics, and CAS Center for Excellence and Synergetic Innovation Center in Quantum Information and Quantum Physics, University of Science and Technology of China, Hefei 230026, P. R. China}

\begin{abstract}
Since the outbreak of COVID-19, many data analysis have been done. Some of them are based on the classical epidemiological approach that assumes an exponential growth, but a few studies report that a power-law scaling may provide a better fitting to the currently available data. Hereby, we examine the epidemic data in China mainland (01/20/2020--02/24/2020) in a log-log scale, and indeed find that the growth closely follows a power-law kinetics over a significantly wide time period. The exponents are $2.48(20)$, $2.21(6)$ and $4.26(12)$ for the number of confirmed infections, deaths and cured cases, respectively, indicating an underlying small-world network structure in the epidemic. While no obvious deviations from the power-law growth can be seen yet for the number of deaths and cured cases, negative deviations have clearly appeared in the infection number, particularly that for the region outside Hubei province. This suggests the beginning of the slowing-down of the spreading due to the huge containment effort. Meanwhile, we find that despite the dramatic difference in magnitudes, the growth kinetics of the infection number exhibits much similarity for Hubei and the region outside Hubei. On this basis, in the log-log plot, we rescale the infection number for the region outside Hubei such that it overlaps as much as possible with the total infection number in China, from which an approximate extrapolation yields the maximum of the epidemic around March 3, 2020, with the number of infections about $83,000$. Further, by analyzing the kinetics of the mortality in the log-log scale, we obtain a rough estimate that near March 3, the death rate of COVID-19 would be about $4.7\%\thicksim 5.0\%$ for Hubei and $0.8\%\thicksim1.0\%$ for the region outside Hubei. We emphasize that our predictions may be quantitatively unreliable, since the data analysis is purely empirical and various assumptions are used.
\end{abstract}


\maketitle

Since December 2019, the novel coronavirus COVID-19 originated in Wuhan have already spread throughout the world, and the total number of confirmed infections has exceeded over 74,000 in China due to the high infectivity of coronavirus COVID-19. In the past week (02/12/2020--02/24/2020), the death rate  has increased from $2.2\%$ to $2.9\%$ for Hubei province and from $0.5\%$ to $0.7\%$ for the other provinces in China (the region outside Hubei). Many analyses have been done for the currently available epidemic data~\cite{Zhou2020,Goetz2020,Ziff2020,Zhang2020,Liang2020,Chen2020,Roosa2020,Wu2020,Yuan2020,Zhao2020,Xiong2020,Hebert-Dufresne2020}, and significant efforts have been spent on forecasting the trends of spreading~\cite{Zhang2020,Liang2020,Chen2020,Roosa2020,Wu2020,Yuan2020,Zhao2020,Xiong2020}.
In the analysis of the empirical data, several studies make use of the classical epidemiological approach that assumes an exponential growth of the disease~\cite{Zhou2020,Goetz2020}. However, Brandenburg suggested that the growth of the disease could be described by a quadratic function~\cite{Goetz2020}, and Ziff {\it et al.} pointed out that a power-law kinetics with fractal exponent $\alpha \approx 2.25$~\cite{Ziff2020} provides a better fit to the current data for the number of deaths. For different regions of China, Maier {\it et al.} found that the number of confirmed infections can be effectively fitted by the pow-law with exponents $2.1\pm0.3$~\cite{Maier2020}. These indicate the emergence of an underlying fractal or small-world network of connections between susceptible and infected individuals, which can be a consequence due to the interplay between the virus spreading and the inhibition like the huge containment endeavor in China. An alarming signal in the study by Ziff {\it et al.}~\cite{Ziff2020} is that no deviations from the power-law growth can be observed yet for the number of deaths, and thus that no prediction can be made for the beginning of the end of the epidemic.

\begin{figure}[h]
\centering
\includegraphics[width=0.7\columnwidth]{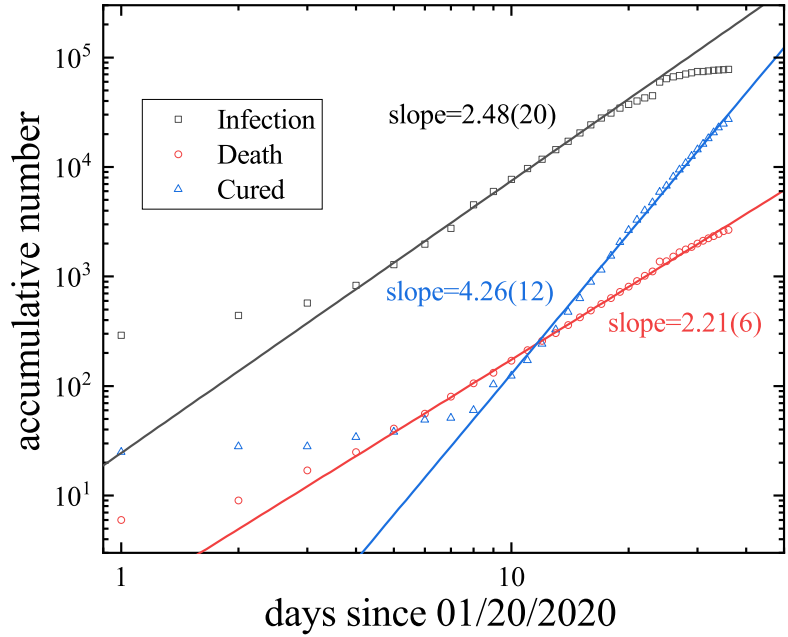}
\caption{The growth of the confirmed infections, cured cases and deaths in China from January 20, 2020 to February 24, 2020. Note that the criterion for the confirmed infections was changed on February 12, 2020, and the number had a jump at that day. The line for the cured cases (deaths) is the fitting result without the early $5$ ($7$) data points and the last $5$ data points, and that for the infections is obtained by fitting the data points from day $3$ to day $18$. Data Source: National Health Commission of the People's Republic of China\cite{g1} and Health Commission of Hubei Province\cite{g2}.}
\label{fig1}
\end{figure}

\begin{figure}[t]
\centering
\includegraphics[width=1.0 \columnwidth]{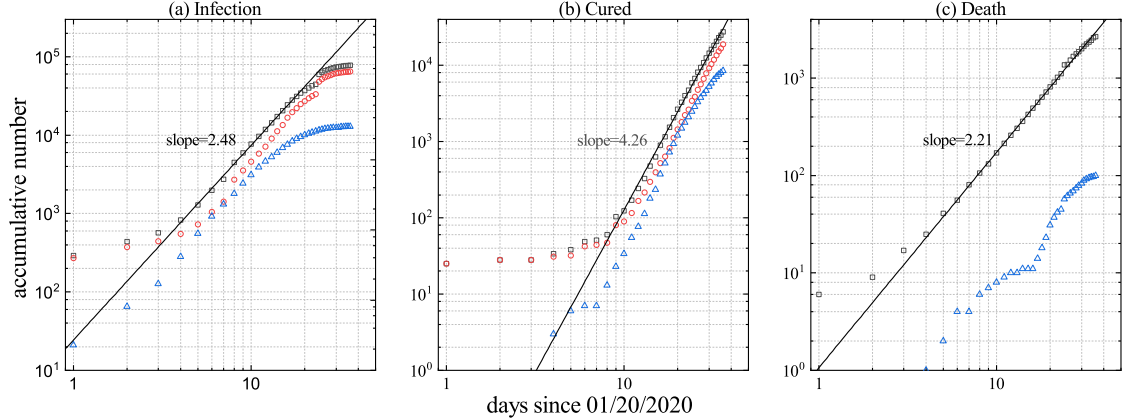}
\caption{The accumulative number of confirmed infections (a), cured cases (b) and deaths (c) in China, in Hubei province, and in the region outside Hubei province from January 20, 2020 to February 24, 2020. Since the numbers of deaths for China and for Hubei province almost coincide with each other, we only draw the number for China in sub-figure (c) for easy visual reference. Data Source: National Health Commission of the People's Republic of China\cite{g1} and Health Commission of Hubei Province\cite{g2}.}
\label{fig2}
\end{figure}

In this work, we reexamine the epidemic data in China mainland starting from January 20, 2020 till today (02/24/2020), and consider the cumulative number of confirmed infections $N_{\rm infec}$, of cured cases $N_{\rm cured}$ and of deaths $N_{\rm death}$. A log-log plot is shown in Fig.~\ref{fig1}, where the beginning of the time period is set as $t=1$ for January 20, 2020. The figure demonstrates that all the three sets of numbers exhibit power-law growing kinetics over a significantly wide time period, and the fractal exponent is $\alpha_{\rm infec} \approx 2.5$ for $N_{\rm infec}$, $\alpha_{\rm death} \approx 2.2$ for $N_{\rm death}$ and $\alpha_{\rm cured} \approx 4.3$ for $N_{\rm cured}$. Apart from the exponent $\alpha_{\rm death} \approx 2.2$, it is confirmed that no obvious deviation can be seen for the power-law growth of $N_{\rm death}$ till 02/22/2020. Nevertheless, negative deviations have clearly appeared in the number of confirmed infections since 02/12/2020 (Fig.~\ref{fig1}). This drop-off tendency is more pronounced in the growth kinetics of $N_{\rm infec}$ for the region outside Hubei, see the blue-triangle data points in Fig.~\ref{fig2}. This suggests that due to the huge containment endeavor in China for the past few weeks, the virus spreading has begun to slow down. Starting from 02/01/2020, the growth of the number of cured cases has speeded up rapidly, with the fractal dimension $\alpha_{\rm cured} \approx 4.3$ significantly larger than the other two exponents $\alpha_{\rm infec} \approx 2.5$ and $\alpha_{\rm death} \approx 2.2$ (Fig.~\ref{fig1}).

For a better understanding of this power-law kinetics and the fractal exponents $\alpha$, one can take into the social contact network, which confines the social contact and thus the spreading of the disease~\cite{Kuperman2001,Riley2015}. Note that the contact network in the classical epidemiological model is fully connected. For a two-dimensional planar network, the infected population can only grow in the periphery of them and the number of infection $N_{\rm infec}$ should grow in a quadratic form $t^2$, as pointed out in Ref.~\cite{Brandenburg2020}. The fractional exponent $\alpha_{\rm infec} \approx 2.5 > 2$ reveals that dense (long-range) connections exist in the underlying network for the spreading of COVID-19, revealing the so-called small-world property~\cite{Watts1998,Watts2004}. From the fact that $\alpha_{\rm cured} \approx 4.3$ is significantly larger than $\alpha_{\rm infec}$, it is also suggested that the network in terms of medical treatments is much more densely connected than that for $N_{\rm infec}$, consistent with the fact that more than $30,000$ doctors and nurses have been sent to Hubei province from the other provinces in China.

At the day of 02/24/2020, about $83\%$ of $77,658$ cumulative confirmed cases are from Hubei province. Thus, we separately plot the number for Hubei and for the region outside Hubei in Fig.~\ref{fig2}. It is observed that the growth kinetics of $N_{\rm infec}$ and $N_{\rm cured}$ exhibit much similarity for Hubei and for the other region, although their magnitudes of the infection numbers and the dates for the starting of the virus spreading can be rather different. This might suggest a similar spreading mechanism for different areas of China. One special case is the growth of the deaths outside Hubei province (Fig.\ref{fig2} (c)), which could because that the number of deaths outside Hubei is too small for statistics. In spite of this, we also find that from 02/12/2020, it started to follow a similar scaling as that for Huber province.

\begin{figure}[t]
\centering
\includegraphics[width=1.0 \columnwidth]{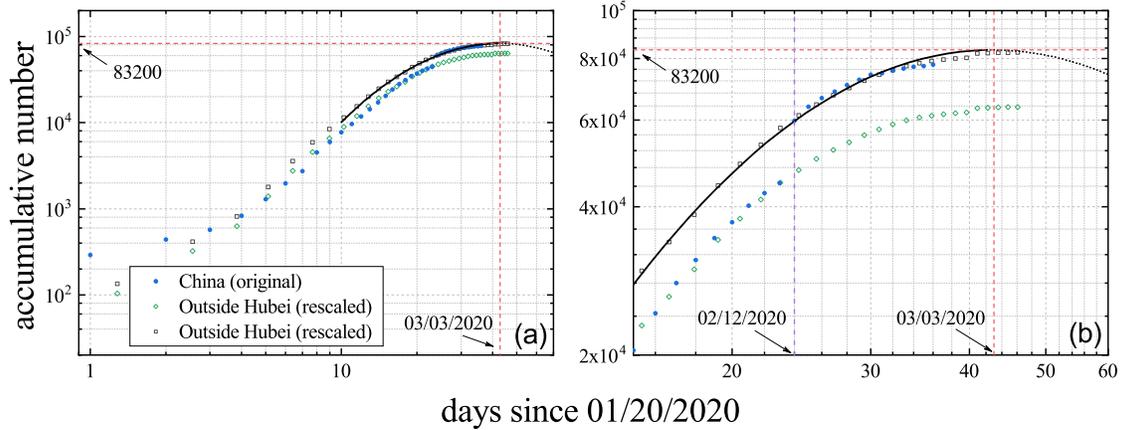}
\caption{The rescaling of the accumulative number of infections for the region outside Hubei. (a) In the log-log plot, we move the data for the region outside Hubei, along both the horizontal and the vertical axis, to overlap with those for China as much as possible. Due to the jump of the accumulative number of infections on February 12, 2020, the data overlapping is done for the time period before February 12, 2020 (green  diamond), as well as after February 12, 2020 (black  square). We fit the rescaled data for the region outside Hubei after 02/12/2020 to ansatz $f(x)=a(x-x_0)^2+y_0$ by gradually excluding the early-stage data points, and obtain $a=-2.30(4)$, $x_0=1.63(1)$ and $y_0=4.92(1)$. The point $(x_0=1.63\pm 0.01, y_0=4.92 \pm 0.01)$, denoted by the crossing point of the red dashed lines, suggests that the maximum number of infections about $83\,000 \pm 20\,00$ would arrive about $43 \pm 2 $ days after January 20, \emph{i.e.}, March 3, 2020. The fitting curve after the extreme point (indicated by dot line) has no physical implication. (b) A zoom-in plot for the data after 02/03/2020.}
\label{fig3}
\end{figure}

With the above observations, we attempt to have a preliminary estimate for the ultimate extent of the epidemic on the basis of the current data. We understand that our estimate can be quantitatively unreliable, since it relies on various factors like the adopted analysis method, the incompleteness of the available information, and the assumption of no new outbreak of COVID-19 \emph{etc}.

In the log-log plot shown in Fig.~\ref{fig2}(a), we try to move the growth kinetics of $N_{\rm infec}$ for the region outside Hubei, along both the horizontal and the vertical axis, such that it overlaps as much as possible with that for China. Note that due to the change of the criteria for infections at the day of 02/12/2020, the number $N_{\rm infec}$ has a jump, and one can choose to achieve the best overlap for the time period before or after 02/12/2020. The results are demonstrated in Fig.~\ref{fig3}(a), and a zoom-in plot for the data after 02/03/2020 is shown in Fig.~\ref{fig3}(b). It seems that the rescaled curve for the period before 02/12/2020 (green diamonds) does not lead to a reasonable extrapolation of the total infection number in China.

We then focus on the rescaled data points in black squares, and apply a polynomial-function fitting for it by gradually excluding the data points for the early stage of $t$. We adopt a simple function as $y=a(x-x_0)^2+y_0$, where $(x_0,y_0)$ specifies the coordinates of the maximum point and $a$ is a free parameter. The least-squares fitting yields an estimate $(x_0=1.63(1), y_0=4.92(1)$, giving $t_0 = 10^{x_0} \approx 43 \pm 2$ (days) and $N_{\rm infec,0}=10^{y_0} \approx 83200 \pm 2000$. In other words, with the assumption of no new break of COVID-19 and of the reliability of such an extrapolation, the epidemic would reach its maximum near March 3, 2020 with an uncertainty of 2 days, and the maximum number of infections in China would be about $ 83\,000 \pm 2,000$. While these fitting results can be quantitatively inaccurate, Fig.~\ref{fig3}(b) seems to suggest that with the current trend, the total number of infections is unlikely to exceed $100,000$.

\begin{figure}[t]
\centering
\includegraphics[width=0.6\columnwidth]{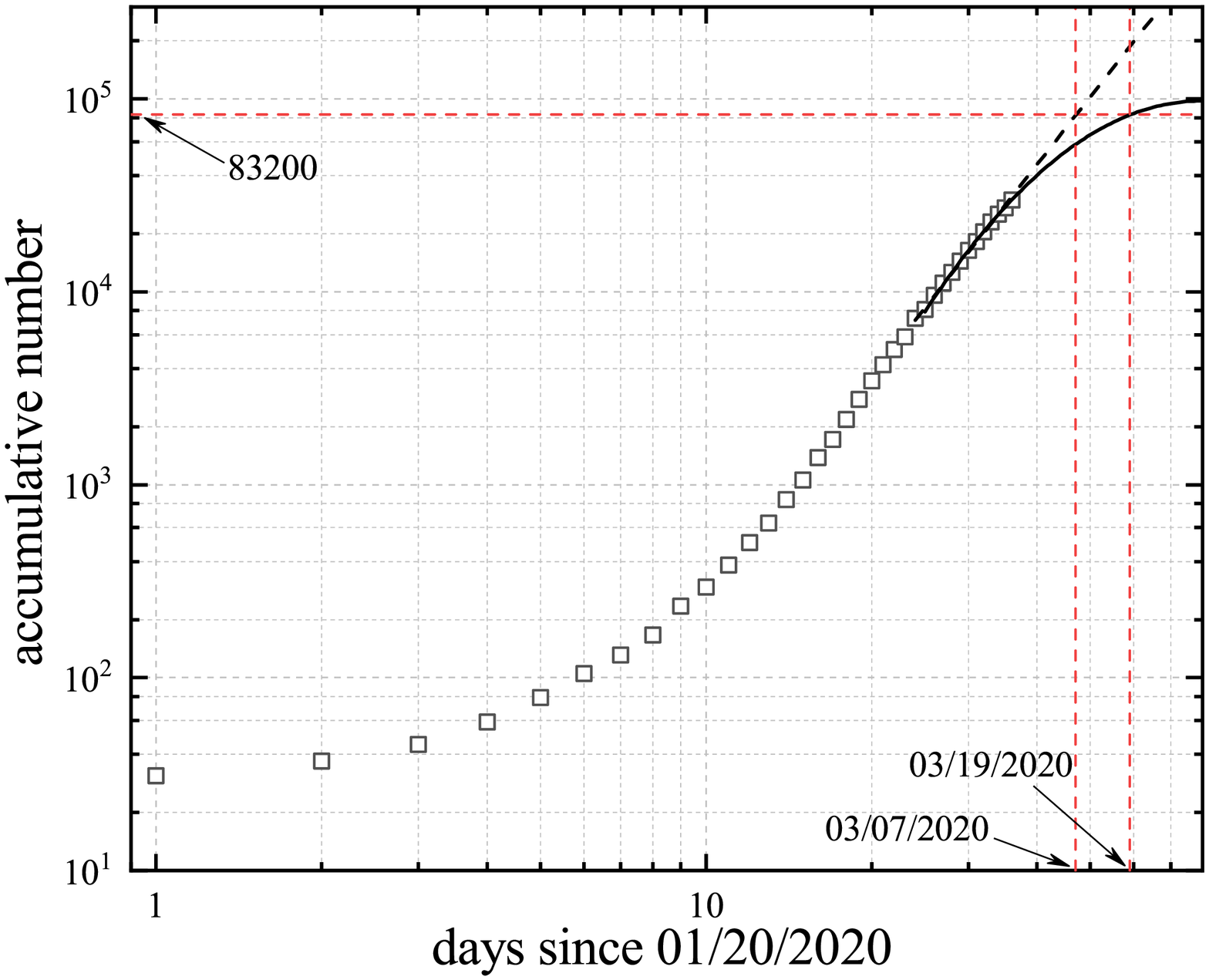}
\caption{The growth of the total number of cured cases and deaths in China. The two fitting functions (in log-log plot) used here are $f(x)=kx+b$ and $f(x)=a(x-x_0)^2+y_0$ with $k=3.63(8)$, $b=-1.15(12)$, $a=-4.33(150)$, $x_0=1.90(16)$ and $y_0=4.99(28)$. The crossing points of the red dashed lines indicate the days that the total number of cured cases and deaths reaches the maximum $83,200$, which is predicted by Fig.\ref{fig3}. This gives an estimation for the end day of the epidemic 03/07/2020-- 03/19/2020.}
\label{fig4}
\end{figure}

\begin{figure}[t]
\centering
\includegraphics[width=0.6\columnwidth]{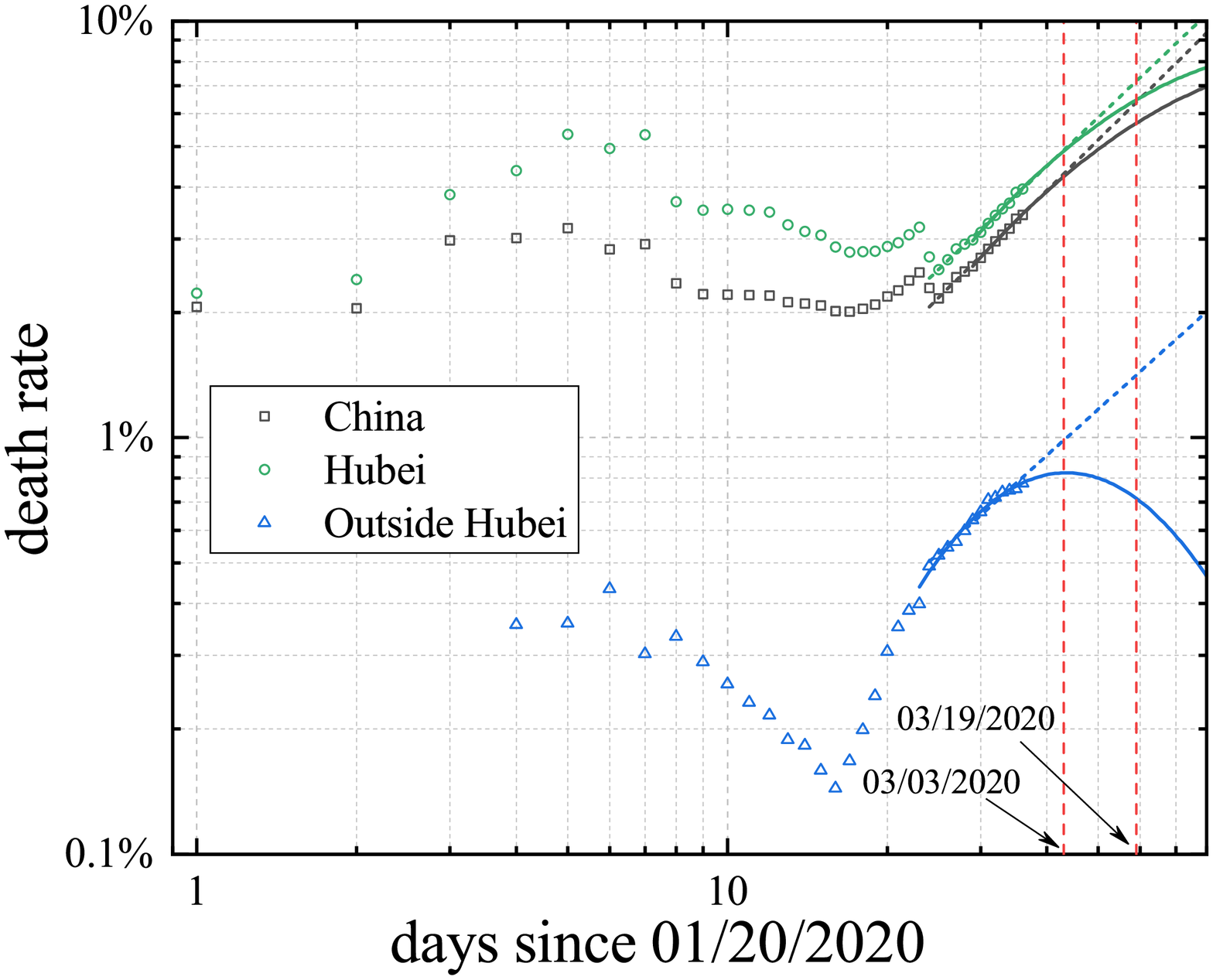}
\caption{The death rate for China, Hubei province and the region outside Hubei province. Here, the death rate is defined as the ratio of the accumulative number of the deaths and that of the confirmed infections. The two fitting functions (in log-log plot) used here are $f(x)=kx+b$ and $f(x)=a(x-x_0)^2+y_0$. For China, $k=1.26(5)$, $b=-3.42(8)$, $a=-1.02(208)$, $x_0=2.16(132)$ and $y_0=-1.09(160)$. For Hubei province, $k=1.21(6)$, $b=-3.29(9)$, $a=-1.15(223)$, $x_0=2.09(111)$ and $y_0=-1.07(150)$. For the region outside Hubei, $k=1.15(12)$, $b=-3.88(20)$, $a=-3.56(204)$, $x_0=1.64(16)$ and $y_0=-2.08(8)$. The red dashed line indicates the day that the number of infections reaches the maximum, \emph{i.e.}, 03/03/2020, and the latest end day, \emph{i.e.}, 03/19/2020.}
\label{fig5}
\end{figure}

We can further fit the data of the total number of cured cases and deaths as shown in Fig.\ref{fig4}. The crossing point of this fitting curve with the estimation of the maximum infections $83,200$ can be thus treated as the end day of the epidemic. As shown in Fig.\ref{fig4}, with the linear function $y=kx+b$ and the quadratic function $y=a (x-x_0)^2+y_0$, the end day is estimated to range from 03/07/2020 to 03/19/2020.

In Fig.\ref{fig5}, we further plot the kinetics of the death rate in the log-log scale, and fit the data after 02/12/2020 by both the linear function $y=kx+b$ and the quadratic function $y=a (x-x_0)^2+y_0$. The extrapolated death rate is then calculated at the day of 03/03/2020, which is about $4.7\%\thicksim 5.0\%$ for Hubei province, $0.8\%\thicksim1.0\%$ for the region outside Hubei and $4 \%\thicksim 4.5\%$ for the whole country of China. For the latest end day estimated in Fig.\ref{fig4}, \emph{i.e.}, 03/19/2020, the corresponding extrapolated death rates are about $6.5\%\thicksim 7.5\%$, $0.7\%\thicksim1.5\%$, and $5.5 \%\thicksim 6.5\%$ for for Hubei province, the region outside Hubei and the whole country of China, respectively.

In summary, we find the growths, for the number of infections, of cured cases and of deaths,  all approximately follow power-law kinetics over a significantly wide time period, indicating the small-world property of the underlying networks. Starting from 02/12/2020, the clear negative deviations from the power-law growth have occurred in the number of infections, reflecting that the spreading of COVID-19 has begun to slow down due to the huge containment endeavor in China. The fractal dimension for the number of cured cases is significantly bigger than those for the number of infections and deaths, suggesting a densely connected small-world network because of enormous medical inputs in China. By a data-collapsing method in the log-log plot, it seems that the epidemic might reach its maximum around March 3, 2020, with the number of infections about $83,200 \pm 2,000$. The mortality rates are analyzed in the log-log plot, and their extrapolated values are calculated for the days of 03/03/2020 and 03/19/2020. However, we emphasize that our estimates may be quantitatively unreliable, since the data analysis is purely empirical and various assumptions are used.

\section*{Acknowledgments}

This work is partly motivated by the private communication with Robert M. Ziff from University of Michigan. YD acknowledges the support by the National Science Foundation of China for Grant No.11625522.

\bibliography{ref}

\end{document}